\documentclass{article}
\usepackage{amsmath, amssymb, verbatim}
\usepackage[dvips]{graphicx}

\title{Towards Skyrmion Stars: Large Baryon Configurations in the 
Einstein-Skyrme Model}
\author{Bernard M. A. G. Piette\footnote{email:b.m.a.g.piette@durham.ac.uk}
\quad and \quad 
Gavin I. Probert\footnote{email:g.i.probert@durham.ac.uk}}

\begin{document}

\maketitle

\begin{center}
                                                                             
{\em Department of Mathematical Sciences,
University of Durham,
Science Laboratories,
South Road,
Durham. DH1 3LE.
UK}
  
\end{center}


\abstract
We investigate the large baryon number sector of the Einstein-Skyrme model as 
a possible model for baryon stars. Gravitating hedgehog skyrmions have been 
investigated previously and the existence of stable solitonic stars excluded 
due to energy considerations\cite{Bizon:1992}. However, in this paper we 
demonstrate that by generating gravitating skyrmions using rational maps, we 
can achieve multi-baryon bound states whilst recovering spherical symmetry in 
the limit where $B$ becomes large.
\newpage


\section{Introduction}
The Skyrme model, in its initial form, was proposed and developed by T.H.R. 
Skyrme in a series of papers as a non-linear field theory of 
pions \cite{Skyrme:1961}, \cite{Skyrme:1962}. Skyrme's initial idea was to 
think of baryons (in particular the nucleons) as secondary structures arising 
from a more fundamental mesonic fluid. The key property of the model was that 
the baryons arose as solitons in a topological manner and thus possessed a 
conserved topological charge identified with the baryon number.

The lowest energy stable solutions of the model are termed 
\textit{Skyrmions} and can be thought of as baryonic solitons. The Skyrme 
model has been very successful in modelling the structures of various nuclei 
and has been shown by Witten et al. \cite{Witten:1983} to possess the general 
features of a low energy effective field theory for QCD.

Some studies of the Skyrme model coupled to gravity have previously been 
undertaken \cite{Bizon:1992}, \cite{Glendenning:1988}, \cite{Volkov:1999}, 
mainly with the motivation of a comparison of its features with those of other 
non-linear field theories coupled to gravity. Of particular note is the 
Einstein-Yang-Mills theory, in which gravitationally bound configurations of 
non-abelian gauge fields are produced.

Other reasons for studying the Einstein-Skyrme model are cosmological and 
astrophysical ones. Various authors have studied black hole formation in the 
model, with the conclusion that the so-called \textit{no-hair} conjecture may 
not hold \cite{Luckock:1986}, \cite{Droz:1991}.

The purpose of this paper is to study large baryon number Skyrmions or 
configurations of Skyrmions in the Einstein-Skyrme model. In particular, we 
wish to investigate if stable solitonic stars could exist within the model and 
to compare their properties to those of neutron stars. 

Preliminary studies of Skyrmion stars have predicted instability to single 
particle decay \cite{Bizon:1992}. However this was done using the 
hedgehog ansatz for baryon number larger than $1$ which is known to lead 
to unstable solutions even for the usual Skyrme model. Since then, it has been 
shown that the Skyrme model has stable shell-like solutions\cite{BS1997} 
which can be well approximated by the so called rational map ansatz 
\cite{Houghton:1998}.

In this paper we use the rational map ansatz and its extension to multiple 
shells to construct configurations in the Gravitating Skyrme model that have
a very large number of baryon. We show that those configurations, contrary to
the hedgehog ansatz are bound even for very large baryon numbers.

To construct configurations that have a baryon 
number comparable to that of neutron star, we have to introduce a further 
approximation, which we call the ramp ansatz. We show that this anstaz 
introduces further errors of only a few percent and we use it to compute
very large Skyrmion configurations.

The paper is organised as follow:
first we outline the Einstein-Skyrme model and discuss the main 
features of the results on static gravitating $SU(2)$ hedgehogs obtained by 
Bizon and Chmaj \cite{Bizon:1992}. We then use the rational map ansatz to 
construct shell like gravitating multibaryon configurations and show that for 
a fixed value of the coupling constant, the configurations exist only when 
the baryon number is below a certain critical value.
Finally we introduce a ramp profile approximation to construct solutions 
with extremely high baryon numbers. We show how accurate it is and use it
to construct Skyrmion stars configuration. 


\section{The Einstein-Skyrme Model}

The action for gravitating Skyrmions is formed from the standard Skyrme action 
for the matter field and the Einstein-Hilbert action for the gravitational 
field.
\begin{equation}
\label{action}
S=\int_{M} \sqrt{-g} \left(\mathcal{L}_{Sk} - \frac{R}{16\pi G} \right) d^4x .
\end{equation}
Here $\mathcal{L}_{Sk}$ is the Lagrangian density for the Skyrme model defined 
on the manifold $M$:
\begin{equation}
\label{Lagrangian}
\mathcal{L}_{Sk} = \frac{F_{\pi} ^2}{16} Tr (\nabla_{\mu} U \nabla^{\mu} U^{-1})
 + \frac{1}{32 e^2} Tr [(\nabla_{\mu} U) U^{-1}, (\nabla_{\nu} U) U^{-1}]^2,
\end{equation}
where $U$ belongs to $SU(2)$.
As we eventually wish to study baryon stars, we take a spherically symmetric 
metric, such as associated with the line element
\begin{equation}
\label{metric}
ds^2 = -A^2 (r)\left( 1 - \frac{2m(r)}{r}\right) dt^2 + 
 \left(1-\frac{2m(r)}{r}\right)^{-1} dr^2 + r^2(d\theta^2 + 
  \sin^2 \theta d\phi^2),
\end{equation}
where $A(r)$ and $m(r)$ are two profile functions that must be determined by 
solving the Einstein equations for the model.
Our choice of ansatz is motivated by the fact that although in some cases 
we will be studying 
non-spherical Skyrmion configurations, the regime we are primarily interested 
in (i.e. Skyrmions of extremely high baryon number) will be shown to admit 
quasi spherical solutions. Also, for realistic values of the couplings,
the gravitational interaction is small compared to the Skyrme interaction and 
thus the use of a spherical metric even with non-spherical configurations, is 
not a great problem.  

From \eqref{metric}, it can be shown that the Ricci scalar is
\begin{equation}
\label{Ricci Scalar}
R=\frac{-2}{Ar^2} \left(-A''r^2 - 2A'r +2A''rm +A'm +3A'rm' + Arm'' + 2Am' 
\right)
\end{equation}
which, after integrating various terms by parts and noting that asymptotic 
flatness requires both $A(r)$ and $m(r)$ to take a constant value at spatial 
infinity, reduces the gravitational part of the action to
\begin{equation}
\label{Grav Action}
S_{gr} = \int A(r)\left(\frac{-m'(r)}{G}\right)dr + \frac{m(\infty)}{G}.
\end{equation}
For what follows, it will be convenient to scale to dimensionless variables by 
defining $x=e F_{\pi}r$ and $\mu(x)=e F_{\pi}m(r)/2$, resulting in one 
dimensionless coupling parameter for the model, $\alpha=\pi F_{\pi} ^{2} G$. 
We note that taking $F_{\pi}=186Mev$ and $G=6.72 \times 10^{-45} Mev^{-2}$, 
then the physical value of the coupling is $\alpha = 7.3 \times 10^{-40}$.

As the Skyrme field is an $SU(2)$ valued scalar field, at any given time one 
can think of it as a map from $\mathbb{R}^{3}$ to the 
$SU(2)$ manifold. Finite energy considerations impose that the field at 
spatial infinity should map to the same point on $SU(2)$, say the identity. 
Thus, one can simply think of the Skyrme field as a map between three-spheres. 
All such maps fall into disjoint homotopy classes characterised by their 
winding number. This winding number is a conserved topological charge because 
no continuous 
deformation of the field and thus no time evolution, can allow transitions 
between homotopy classes. It is this topological charge that is interpreted 
as the baryon number.


\section{Gravitating Hedgehog Skyrmions}
Gravitating Skyrmions were first studied by Bizon and Chmaj\cite{Bizon:1992} 
who analysed the properties of static spherically 
symmetric gravitating $SU(2)$ skyrmions. Taking the 
\textit{Hedgehog Ansatz} for the Skyrme field
\begin{equation}
U=exp(i \overrightarrow{\sigma}.\hat{r} F(r))
\end{equation}
subject to the boundary conditions
\begin{eqnarray}
\label{Hedge BCs}
F(r=0) &=& \mathcal{B} \pi \\
F(r=\infty) &=& 0
\end{eqnarray}
where $\mathcal{B}$ is the Baryon number associated with the Skyrmion
configuration, they derived the Euler-Lagrange equation for the profiles
$F(r)$, $(A(r)$ and $m(r)$ and found that the model 
admit two branches of global solitonic solutions at each given baryon number, 
which annihilate at a critical value of the coupling parameter. Above 
$\alpha_{crit}$ no further solutions were found. In particular the value of 
the critical coupling decreased quite considerably with increasing baryon 
number as $\alpha_{crit} \approx 0.040378/\mathcal{B}^2$.
It appears that the existence of a critical coupling does not signal the 
collapse of a Skyrmion to form a black hole. In fact the metric factor 
$S(x)=(1-\frac{2\mu(x)}{x})$ is non-zero at $\alpha_{crit}$; 
there simply ceases to be any stationary points of the action above the 
critical coupling.

The major problem with the ansatz \eqref{Hedge BCs} is that it leads to 
unstable solutions, {\it i.e.} for any given value of $\alpha$,
$M_{ADM} (\mathcal{B}=N) > N M_{ADM}(\mathcal{B}=1)$. This is actually the case 
for the pure Skyrme model as well where the hedgehog anstaz \eqref{Hedge BCs} 
with $B>1$ does not correspond to the lowest energy solution for the model. 
The solutions of the pure Skyrme model when $\mathcal{B}>1$ are known not 
to be spherically 
symmetric\cite{Battye:2001} but are stable {\it i.e.} 
$E(\mathcal{B}=N) < N*E(\mathcal{B}=1)$.

It was actually shown by Houghton et al \cite{Houghton:1998}, 
\cite{Manton:2000} that the multi-baryon solutions of the pure Skyrme model
can be well approximated by the so called rational maps ansatz which 
is a generalisation of the hedgehog ansatz. While not radially symmetric, the
ansatz separates its radial and angular dependence through a profile function 
and a rational map respectively.

In the following sections we will generalise the construction of Houghton et 
al to approximate the solution of the Einstein-Skyrme model.

\section{The Rational Map Ansatz}

The rational map ansatz introduced by Houghton 
\textit{et al.}\cite{Houghton:1998} works by decomposing the field 
into angular and radial parts. Using the polar coordinates in $\mathbb{R}^{3}$ 
and defining the 
stereographic coordinates $z=\tan(\theta/2) \exp^{i\phi}$
the ansatz reads 
\cite{Houghton:1998}
\begin{equation}
\label{ratmap}
U = exp\left(i\vec{\sigma}\cdot\hat{n_{R}}F(r,t)\right)
\end{equation}
where 
\begin{equation}
\label{Rat vec}
\hat{n_{R}} = \frac{1}{1+\vert R \vert^{2}}\left(2\Re(R),2\Im(R),1-
              \vert R \vert^{2}\right)
\end{equation}
is a unit vector where $R$ is a rational function of $z$.

It can be shown that the baryon number for Skyrmions constructed in this way, 
is equal to the degree of the rational map providing we take the boundary 
conditions
\begin{eqnarray}
\label{Rat BCs}
F(r=0)&=&\pi\nonumber \\
F(r=\infty)&=&0.
\end{eqnarray}

Substituting the ansatz \eqref{ratmap} into the action for the model and 
scaling to dimensionless variables as earlier, we obtain the following reduced 
Hamiltonian
\begin{eqnarray}
\label{Rat Ham}
H = \frac{16 \pi F_{\pi}}{e} \left[ \displaystyle{\int_0^\infty} A(x) 
 \left( \frac{1}{2} S(x)F(x)'^{2}x^{2} \right. \right. &+& 
  \mathcal{B}sin^{2}F(x)(1 + S(x)F(x)'^{2}) \nonumber \\
&+& \left. \left. \frac{\mathcal{I} sin^{4}F(x)}{2x^{2}} - 
  \frac{\mu'(x)}{\alpha}\right)dx + \frac{\mu(\infty)}{\alpha} \right]
\end{eqnarray}
where
\begin{equation}
S(x) = 1 - {2 \mu(x)\over x}
\end{equation}

From which one obtains the following field equations

\begin{equation}
\label{mueom}
\mu (x) ' = \alpha \left(\frac{1}{2}S(x) x^2 F(x)'^2 + \mathcal{B} \sin^2 F(x) 
            + S(x)\mathcal{B} F(x)'^2 \sin^2 F(x) 
            + \frac{\mathcal{I}\sin^4 F(x)}{2 x^2}\right)
\end{equation}

\begin{eqnarray}
\label{Feom}
F(x)''= & &\frac{1}{S(x)V(x)}\left[ \sin2F(x) \left(\mathcal{B} 
          + S(x)\mathcal{B}F(x)'^2 
          + \frac{\mathcal{I} \sin^2 F(x)}{x^2} \right) \right.\nonumber \\
 &-& \frac{\alpha S(x) F(x)'^3 V(x)^2}{x} 
    - \left. S(x)'F(x)'V(x) -S(x)F(x)'V(x)' \right]
\end{eqnarray}

and
\begin{equation}
\label{Aeom}
A(x)' = \alpha A(x) F(x)'^2 \left( x + 2  \mathcal{B} 
       \frac{\sin^2 F(x)}{x} \right)
\end{equation}

where, for convenience, we have defined $V(x)$ as
\begin{equation}
V(x)=x^2  + 2 \mathcal{B} \sin^2 F(x).
\end{equation}

$\mathcal{B}$ is the baryon number and 
\begin{equation}
\label{I} 
\mathcal{I}=\frac{1}{4\pi} \int \left(\frac{1+|z|^2}{1+|R|^2} \left| 
        \frac{dR}{dz}\right|\right)^4 \frac{2i dz d\overline{z}}{(1+|z|^2)^2}
\end{equation}
Its value depends on the chosen rational map $R$. To compute low energy 
configurations for a given baryon charge $\mathcal{B}$ one must find the 
rational map $R$ or degree $\mathcal{B}$ that minimize $\mathcal{I}$. This has
been done in \cite{Houghton:1998} and \cite{Battye:2001} for several 
values of $\mathcal{B}$. Moreover when $b$ is large, 
one can use the approximation\cite{Battye:2001} 
$\mathcal{I} \approx 1.28 \mathcal{B}^{2}$. The value of $\mathcal{I}$ so 
obtained is then used as a parameter and one can solve equations 
\eqref{mueom} - \eqref{Aeom} for the radial profiles $F(x)$, $A(x)$ and 
$\mu(x)$.

We should point out here that for the pure Skyrme model the rational map 
ansatz produce very good approximation to the multi skyrmion solutions
\cite{Houghton:1998}: the energies are only 3 or 4 percent higher and the
energy densities exhibit the same symmetries and differ by very little.
All the solutions computed by Battye and Sutcliffe\cite{Battye:2001}, 
when $\mathcal{B}$ is not too small, have somehow the shape of a hollow shell. 
The baryon density is very small everywhere outside the shell, while on the 
shell itself, it forms a lattice of hexagons and pentagons.

\section{Hollow Skyrmion Shells}
Using the rational map ansatz, we will now solves the field equations 
\eqref{mueom} - \eqref{Aeom} to compute some low action configurations. These 
solutions will correspond, initially, to a hollow shell of Skyrmions similar 
to the configuration obtained with the rational map anstaz for the pure Skyrme 
model. In the following sections we will show how our ansatz can be 
generalised to allow for more realistic configuration made out of 
embedded shells.

\begin{center}
\begin{figure}[!ht]
\includegraphics[height=8cm,width=14cm, angle=0]{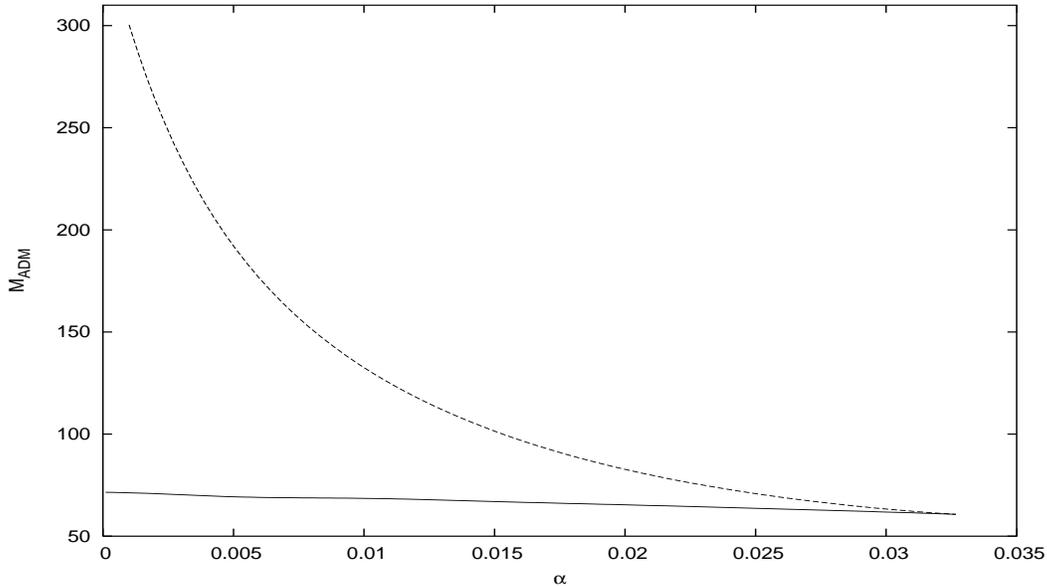}
\caption{\emph{Plot of the two branches of solutions found for B=2 
configurations generated with the rational map ansatz.}}
\label{NumBranches}
\end{figure}
\end{center}

\begin{center}
\begin{figure}[!ht]
\includegraphics[height=8cm,width=14cm, angle=0]{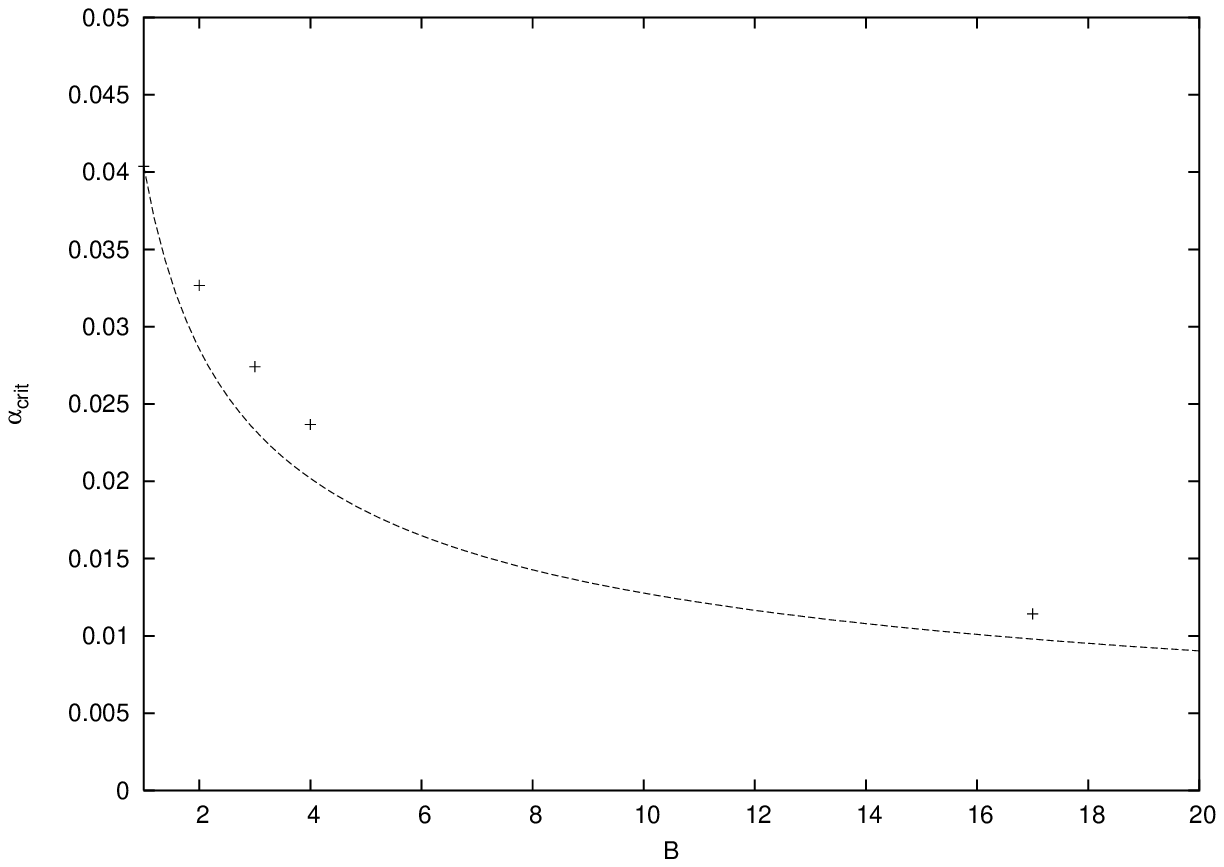}
\caption{\emph{Plot of the decrease in $\alpha_{crit}$ with increasing baryon 
number, for configurations generated with the rational map ansatz. $+$:
$\alpha_cr$ for the minimum value of $I$; curve: 
$\alpha_cr = 0.0404 \mathcal{B}^{-1/2}$}.}
\label{NumCrit}
\end{figure}
\end{center}

The first thing to note about our solutions is that we again obtain two 
branches of solutions at each baryon number (Fig. \ref{NumBranches}). 
Obtaining this same qualitative behaviour is not surprising when one considers 
that the $\mathcal{B}=1$ rational map Skyrmion reproduces the usual 
$\mathcal{B}=1$ hedgehog.
However, the behaviour of the critical coupling itself is drastically altered 
for the rational map generated configurations. Namely, we observe that it 
decreases as approximately $0.040378/B^{\frac{1}{2}}$ (Fig. \ref{NumCrit}). 
In particular this means that for a given value of the coupling, the rational 
map generated skyrmions can possess a much higher topological charge than 
their hedgehog counterparts, before there ceases to be any solutions. 
Quantitatively if $\mathcal{B}_{hedgehog}$ is the maximum baryon number for 
which hedgehog solutions can be found at a given value of the coupling, then 
the highest baryon number rational map solution found at the same value of 
$\alpha$ will be approximately $\mathcal{B}_{hedgehog} ^4$.
Again we observe that the metric function $S(x)$ is non-zero at the critical 
coupling for all the solutions we have found and as such a horizon has not 
formed.

In Table \ref{1shellNum} we present the radius, ADM mass per baryon 
and minimum value 
of the metric function, $S(x)$, for configurations up to the maximum baryon 
number allowed at $\alpha = 1 \times 10^{-6}$. These values were obtained by 
direct numerical solution of equations \eqref{mueom} - \eqref{Aeom}, where we 
have used the boundary data as specified in \eqref{Rat BCs}. 

We didn't didn't use the physical value of $\alpha$ ($7.3  \times 10^{-40}$)
because for this value, the ratio between the width of the shell and its 
radius is so small when we reach the maximum value of $\mathcal{B}$ that it 
becomes very difficult to solve the equation reliably. 
The value  $\alpha = 1 \times 10^{-6}$ is small enough to allow for a
shell with a large baryon number to exist but large enough to make it possible
to compute these solution nears the critical value of $\mathcal{B}$ 
for a single shell configuration.

The major difference between these configuration and the solutions of 
Bizon and Chmadj is that the rational map ansatz configurations 
become more bound when the baryon number increases  
This suggests the possibility that giant gravitating Skyrmions 
can be bound and consequently, that the Skyrme model can be used to 
study baryon stars. 

Another interesting feature of the data is the observed change in the radius 
of the solutions with increasing baryon number. We note that the radius grows 
as approximately $\mathcal{B}^{\frac{1}{2}}$. However there are two main 
deviations from this. Firstly, the constant of proportionality relating the 
radius to the square root of the baryon number decreases slightly but 
persistently as we increase the baryon number, indicating the gravitational 
interaction becoming more important as the number of baryons increases.

As we approach the maximum baryon charge that can exist at 
$\alpha = 1 \times 10^{-6}$, we also notice that the radius of the skyrmion 
actually decreases as we add more baryons. This shows that the gravitation 
pull plays a crucial role near the critical value of the skyrmion.
This is a tantalising property 
when one considers that generally a neutron star's radius must decrease for 
an increase in mass in order to achieve sufficient degenerate neutron pressure 
to support the star.

\begin{table}
\begin{center}
\begin{tabular}{|c|c|c|c|c|}
\hline
$\mathcal{B}$ & $R (\frac{2}{F_{\pi}{e}})$ & $M_{ADM} 
(\frac{F_{\pi}}{2 e top conv}) $ & $S_{min}$ \\
\hline
1 & 0.8763 & 1.2315 & 1.0000 \\
\hline
4 & 1.7728 & 1.1365 & 1.0000 \\
\hline
8 & 2.5065 & 1.1180 & 1.0000 \\
\hline
100 & 8.6829 & 1.0845 & 0.9999 \\
\hline
500 & 19.3994 & 1.0827 & 0.9998 \\
\hline
$1 \times 10^3$ & 27.4314 & 1.0825 & 0.9997 \\
\hline
$1 \times 10^4$ & 86.7192 & 1.0821 & 0.9989 \\
\hline
$1 \times 10^5$ & 274.0397 & 1.0814 & 0.9963 \\
\hline
$1 \times 10^6$ & 864.6968 & 1.0792 & 0.9883 \\ 
\hline
$1 \times 10^7$ & 2715.0729 & 1.0722 & 0.9628 \\
\hline
$1 \times 10^8$ & 8377.4601 & 1.0500 & 0.88192 \\
\hline
$1 \times 10^9$ & 23585.5315 & 0.9743 & 0.6107 \\
\hline
$1.5 \times 10^9$ & 26860.2040 & 0.9463 & 0.5020 \\
\hline
$1.8 \times 10^9$ & 27470.2449 & 0.9302 & 0.4256 \\
\hline
$1.81 \times 10^9$ & 27456.5804 & 0.9296 & 0.4225 \\
\hline
$1.85 \times 10^9$ & 27357.9201 & 0.9274 & 0.4090 \\
\hline
$1.9 \times 10^9$ & 27078.6014 & 0.9246 & 0.3886 \\
\hline
$1.95 \times 10^9$ & 26126.5508 & 0.9217 & 0.3517 \\
\hline
$1.951 \times 10^9$ & 26050.7695 & 0.9217 & 0.3495 \\
\hline
$1.952 \times 10^9$ & 25937.4210 & 0.9216 & 0.3463 \\
\hline
\end{tabular}
\caption{\emph{Properties of the one shell low energy configuration for 
$\alpha = 1 \times 10{-6}$}}
\label{1shellNum}
\end{center}
\end{table}

\begin{figure}[!ht]
\begin{center}
\includegraphics[height=5cm,width=6.5cm, angle=0]{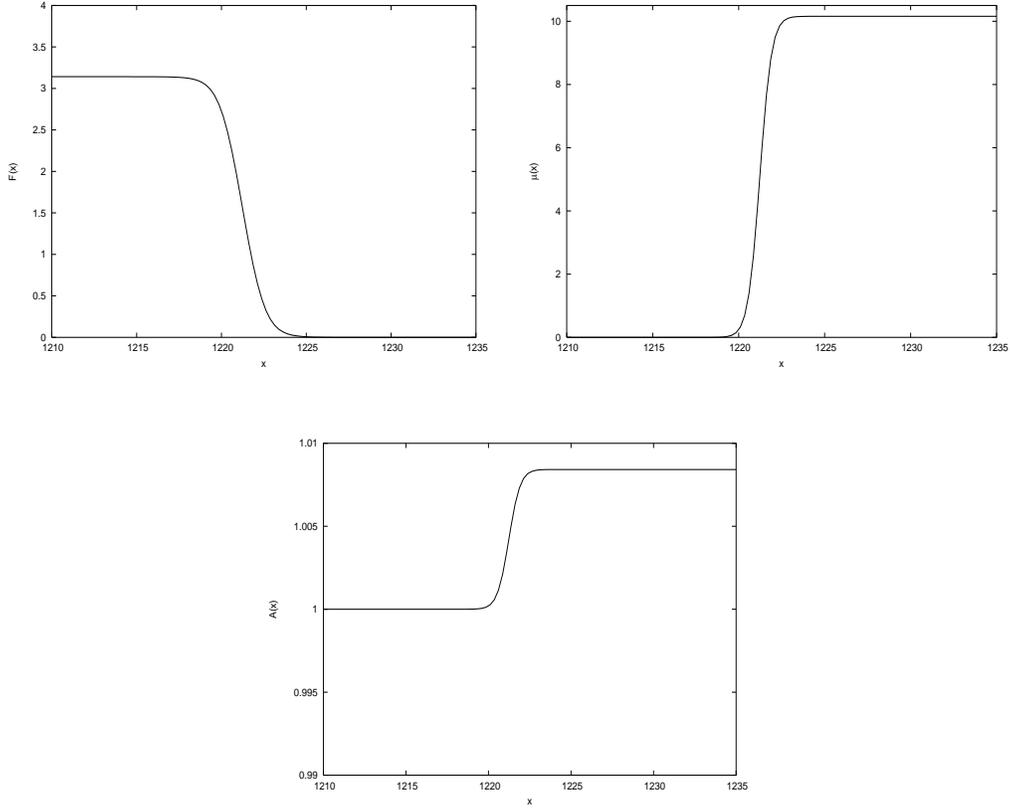}
\caption{Numerical solutions for the profiles $F(x)$, $\mu(x)$ and $A(x)$ when
$\mathcal{B}=2 \times 10^6$ and $\alpha=1 \times 10^{-6}$.}
\label{Num1step}   
\end{center}
\end{figure}

To motivate the further approximation that we will introduce in the next 
section, we now look at the profiles of the configuration that we have 
computed. First of all, we observe 
that the profile function $F(x)$ stays approximately at its boundary value, 
$\pi$, for a finite radial distance before decreasing monotonically over 
some small region and finally attaining its second boundary value, $0$. A 
similar behaviour is seen for both the mass field $\mu(x)$ and the metric 
field $A(x)$ (see Fig. \ref{Num1step}). Furthermore, as we increase the baryon 
number the structure becomes more pronounced, with the distance before the 
fields change (shell radius) increasing significantly, whilst the distance 
over which the fields change (shell width) settles to a constant size. We 
conclude that at large baryon numbers, those configurations correspond 
to hollow shells where the baryons are distributed on a tight lattice over 
the shell. 
As such the, structures are nearly spherical, validating our choice of 
radial metric.

\begin{figure}[!ht]
\begin{center}
\begin{tabular}{cc}
{\includegraphics[height=5cm,width=6.5cm, angle=0]{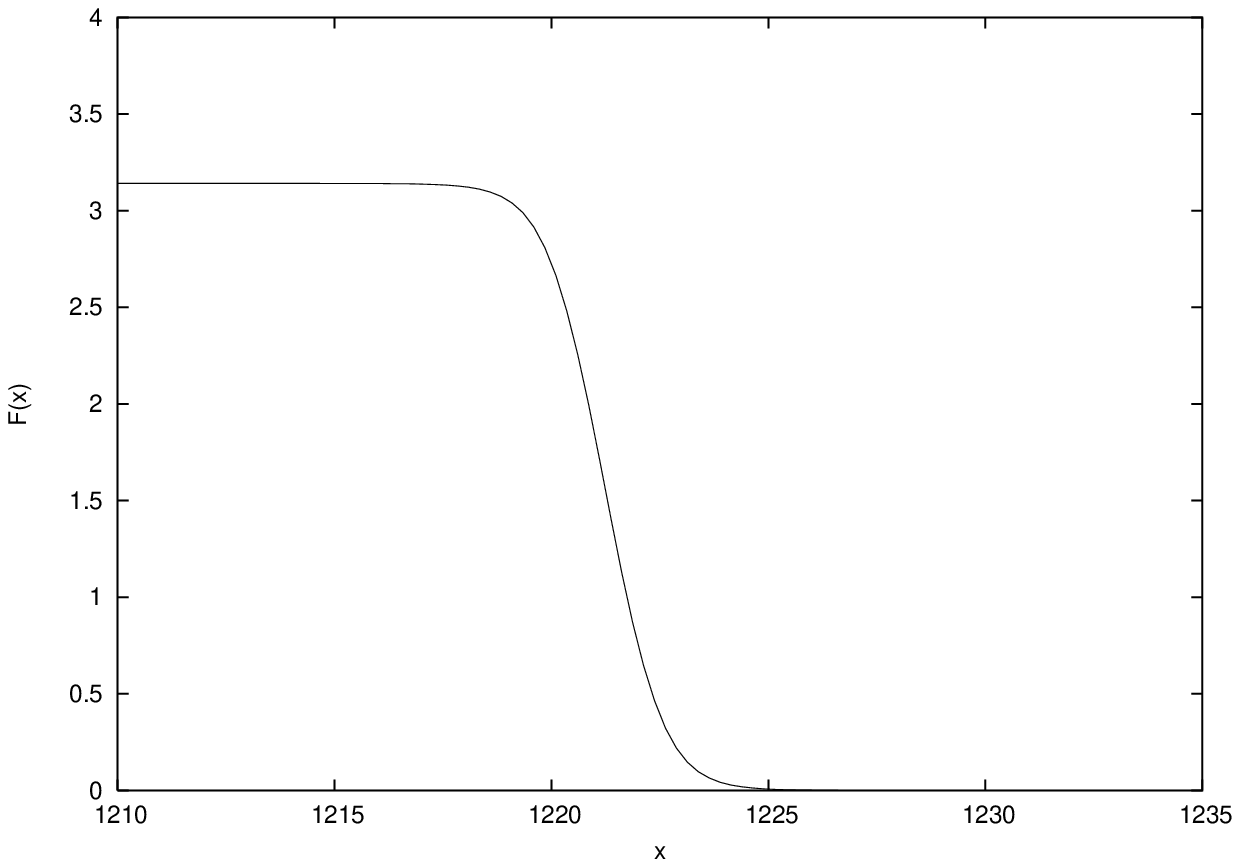}}

& 

{\includegraphics[height=5cm,width=6.5cm, angle=0]{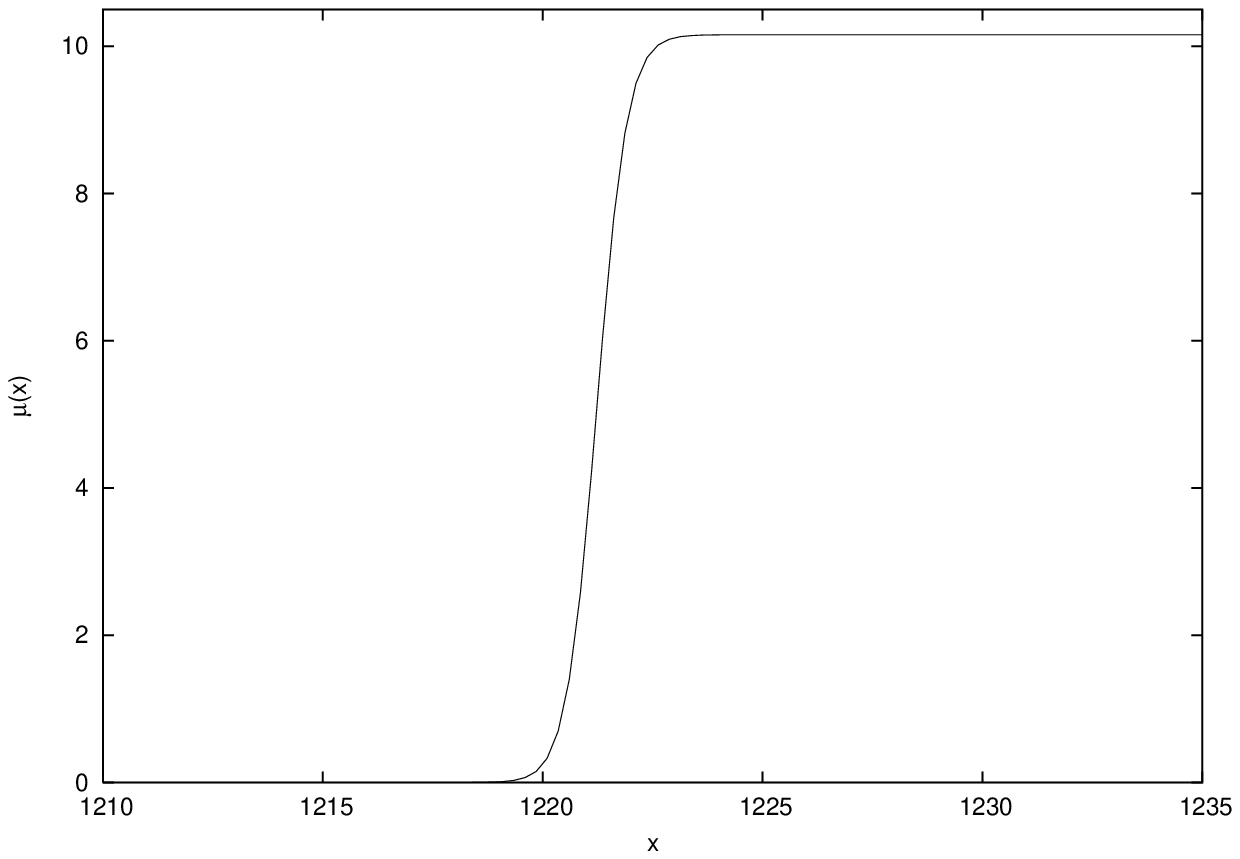}}

\end{tabular}
\end{center}
\end{figure}

Such structures immediately pose an interesting question. Can the gravitating 
Skyrmions exist as shells with more than one layer? To investigate this we 
note that it is possible to modify the boundary condition \eqref{Rat BCs}
to read 
\begin{eqnarray}
F(r=0) &=& N \pi \\
F(r=\infty) &=& 0
\label{Nlayer BCs}
\end{eqnarray}
whilst still ensuring that the Skyrme field is well defined at the origin.
This idea was first used in\cite{Manton:2000} to construct two shell
configurations for the pure Skyrme model. 

\begin{figure}[!ht]
\begin{center}
\begin{tabular}{cc}
{\includegraphics[height=4.8cm,width=6.5cm, angle=0]{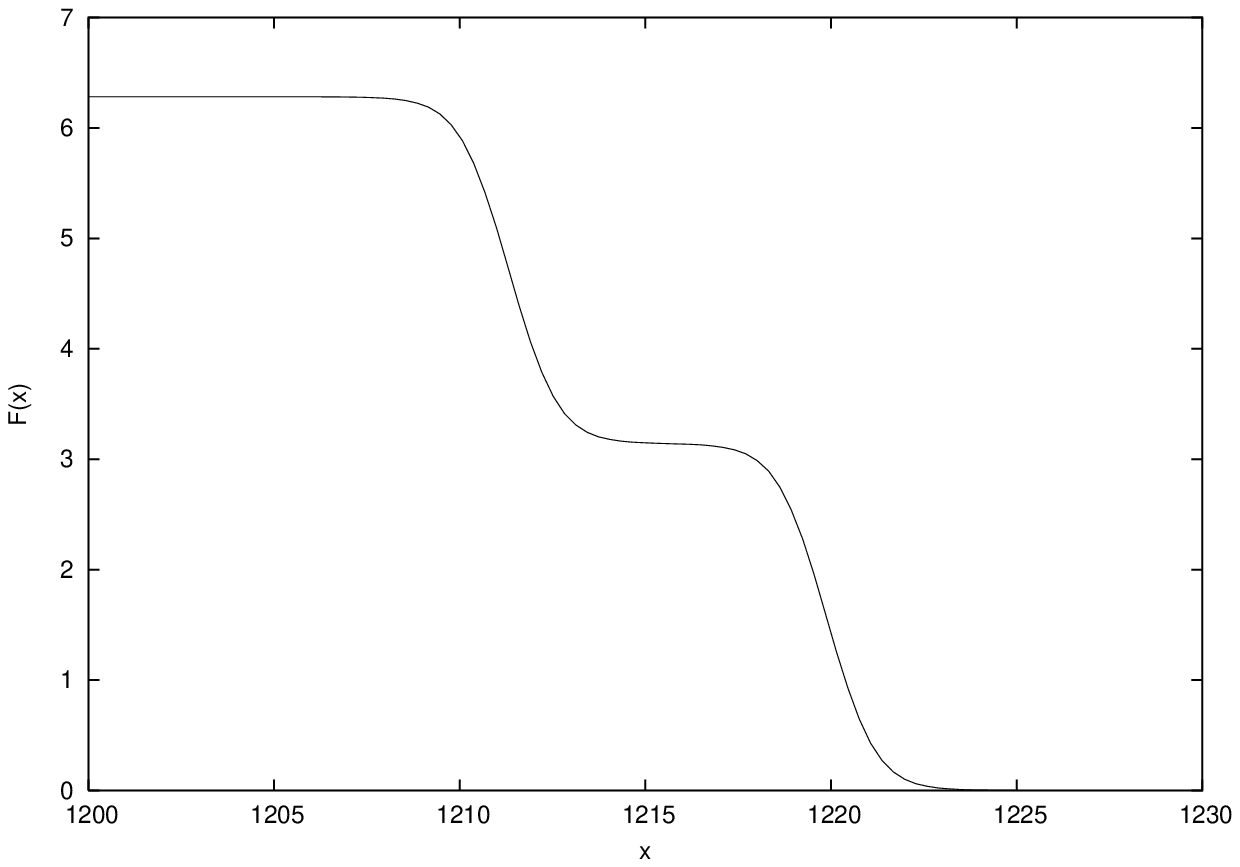}}
& 
{\includegraphics[height=4.8cm,width=6.5cm, angle=0]{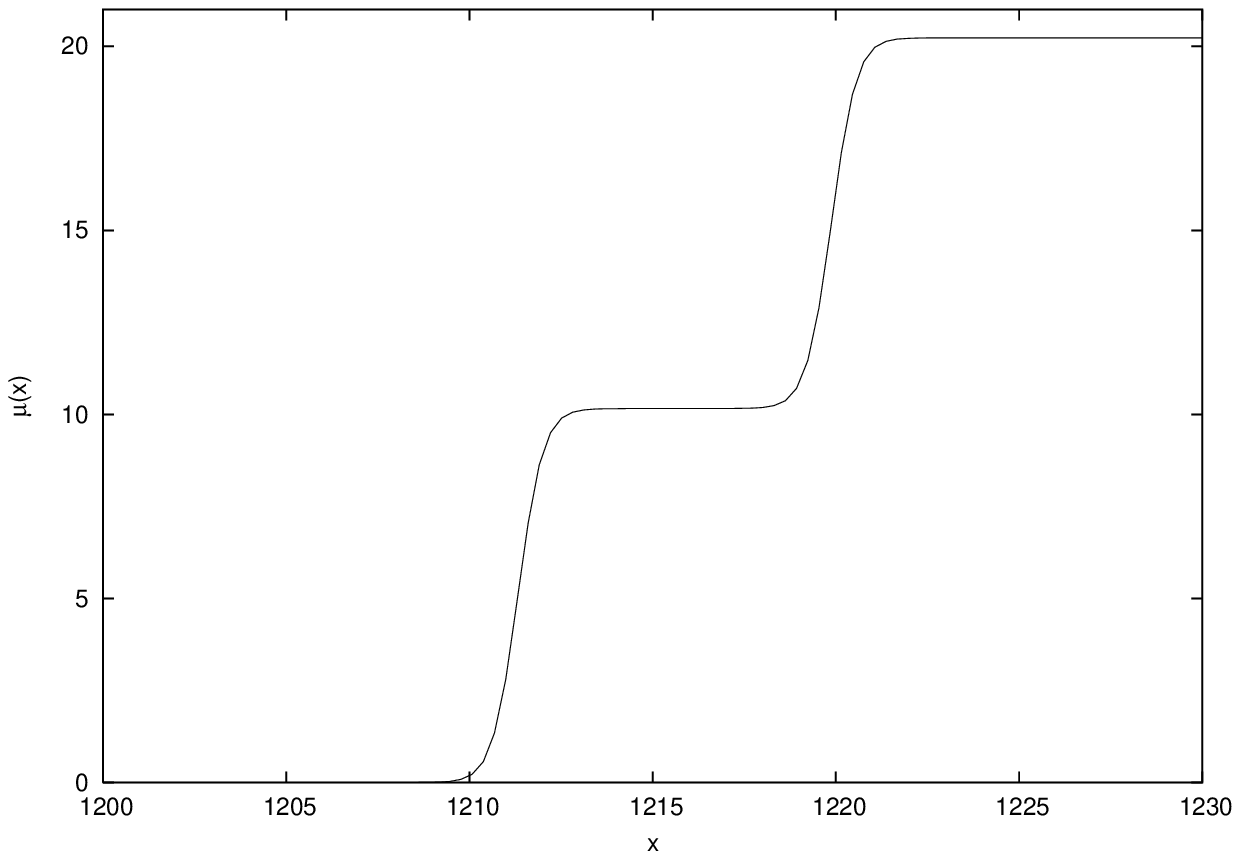}}
\end{tabular}
\end{center}
\end{figure}

\begin{figure}[!ht]
\begin{center}
\includegraphics[height=4.8cm,width=6.5cm, angle=0]{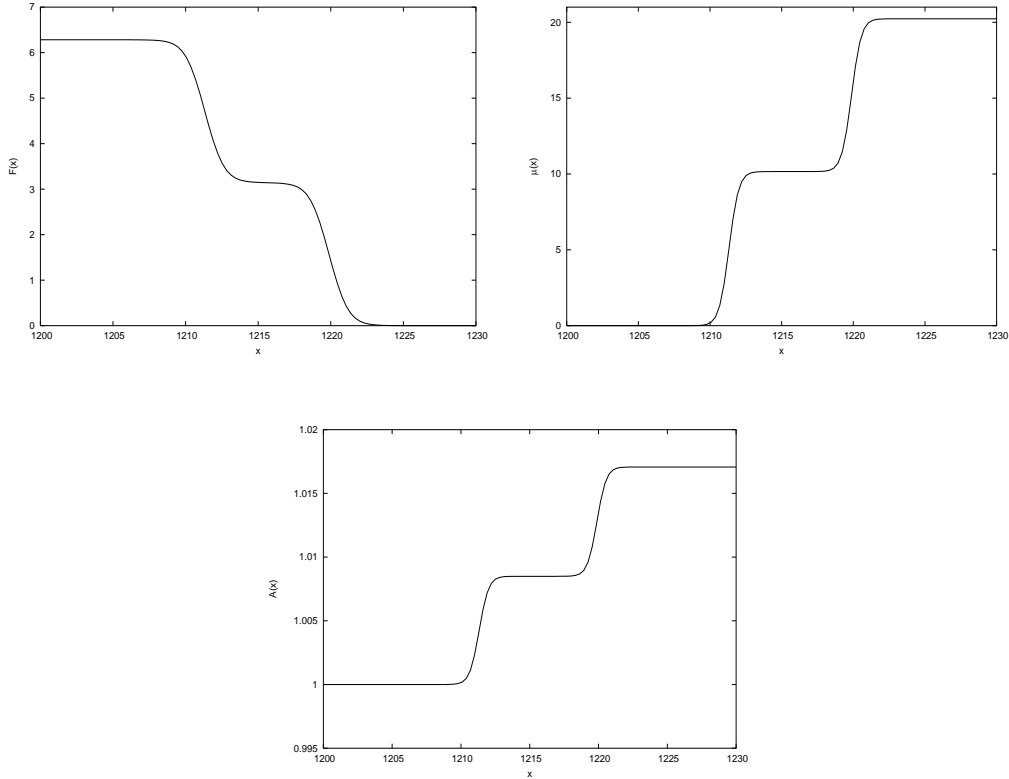}
\caption{Numerical solutions for  for the profiles $F(x)$, $\mu(x)$ and $A(x)$ 
for 2 layers configurations ($F(0)=2\pi$) when $\mathcal{B}=2 \times 10^6$ and 
$\alpha=1 \times 10^{-6}$.}
\label{Num2step}   
\end{center}
\end{figure}

The baryon charge is now $N$ times the degree of the rational map. 
Fig. \ref{Num2step} shows the structure of the solutions we find in this 
case when $N=2$. 
They suggest that the Skyrmion now exists as a $N$-layered structure. This is 
exhibited in the form of the profile, mass and metric functions which 
interpolate between the boundary values in $N$ distinct steps of equal size 
stacked next to each other.

We can therefore think of this as a naive way of constructing a gravitating 
Skyrmion. Instead of using the boundary conditions as in \eqref{Rat BCs} and 
a rational map of degree $\mathcal{B}$ we consider constructing the 
$\mathcal{B}$-Skyrmion using a rational map of degree $\mathcal{B} / N$ 
(with the associated value of $\mathcal{I}$) and the boundary condition 
\eqref{Nlayer BCs}. This is a crude construction as we are effectively 
considering $N$ adjacent shells of baryons, all with the same baryon number. 
We might realistically expect that the baryon number per shell and 
distribution of shells may vary significantly for the minimum energy 
configuration. Nevertheless we shall study the properties of such structures. 
In fact, in the case where the baryon number is large and the number of shells 
is small, we expect this crude construction to be quite valid. That is, we do 
not expect the baryon number to change significantly over the few shells at 
large radius. For the remainder of this section we will restrict ourselves to 
the case where $N = 2$.

Table \ref{2shellNum} summarises the properties of double layered gravitating 
skyrmions up to the maximum baryon charge allowed at 
$\alpha = 1 \times 10^{-6}$. Briefly, we note the main features. Firstly, for 
all baryon numbers, the radius of the double layered solutions is 
significantly less than their single layered counterparts. This is not 
surprising as the baryon charge exists over a thicker region and so the mean 
radius can decrease with the baryon density remaining the same. 
Secondly, when $\mathcal{B}$ is large enough, {\it i.e.} when the double layer
starts to make sense, the double layer solutions are energetically favourable
when one compares the ADM mass with the single layer solutions. 
Finally we note 
that the maximum baryon number allowed (at the given coupling) is almost twice 
as much in the case of the single layer skyrmions.    

\begin{table}
\begin{center}
\begin{tabular}{|c|c|c|c|}
\hline
$\mathcal{B}$ & $R (\frac{2}{F_{\pi}{e}})$ & 
$M_{ADM} (\frac{F_{\pi}}{2 e top conv}) $ & $S_{min}$ \\
\hline
4 & 1.2898 & 1.6179 & 1.0000 \\
\hline
8 & 1.7858 & 1.4072 & 1.0000 \\
\hline
100 & 6.1754 & 1.1363 & 0.9999 \\
\hline
$1 \times 10^3$ & 19.4157 & 1.0913 & 0.9996 \\
\hline
$1 \times 10^4$ & 61.3207 & 1.0833 & 0.9985 \\
\hline 
$1 \times 10^5$ & 193.7006 & 1.0812 & 0.9949 \\
\hline
$1 \times 10^6$ & 610.6271 & 1.0779 & 0.9835 \\
\hline
$1 \times 10^7$ & 1911.3704 & 1.0680 & 0.9475 \\
\hline
$1 \times 10^8$ & 5825.2626 & 1.0362 & 0.8325 \\
\hline
$9.0 \times 10^8$ & 13736.9982 & 0.9302 & 0.4258 \\
\hline
$9.7 \times 10^8$ & 13263.0853 & 0.9224 & 0.3644 \\
\hline
$9.76 \times 10^8$ & 12998.2817 & 0.9217 & 0.3480 \\
\hline
$9.764 \times 10^8$ & 12931.5189 & 0.9216 & 0.3444 \\
\hline
$9.7647 \times 10^8$ & 12895.6984 & 0.9216 & 0.3425 \\
\hline
$9.76472 \times 10^8$ & 12891.4247 & 0.9216 & 0.3423 \\
\hline
$9.764724 \times 10^8$ & 12889.8645 & 0.9216 & 0.3422 \\
\hline
\end{tabular}
\caption{\emph{Table of properties of double layer solutions obtained 
numerically at $\alpha = 1 \times 10{-6}$}}
\label{2shellNum}
\end{center}
\end{table}

Of course the results of this section are not really the main regime of 
interest. We clearly need to study configurations of extremely high baryon 
number (of order $10^{58}$) relevant for baryon stars. 

We will now discuss this high baryon number regime.

\section{The Ramp-profile Approximation}

Unfortunately, at very high baryon numbers, eqns. (\eqref{mueom} - 
\eqref{Aeom}) become difficult to handle numerically. This is largely because 
the radius of the solutions becomes much larger than the distance over which 
the fields change. That is, we need to integrate over a region which is much 
less than $10^{-16} radius$, and so even double precision data types 
have insufficient precision.

Moreover, single shell configurations are not physically relevant and multiple 
shells will only yield configuration that looks like a star if the number of 
layers is very large, typically well over $10^{17}$. With such a large number 
of layers we won't be able to solve the equation numerically as we will need 
at least 10 times as many sampling points for the profile functions.
We must thus resort to another level of approximation: approximate the profile
functions by profiles that are piecewise linear. 
This is inspired by the work of Kopeliovich 
\cite{Kopeliovich:2001} \cite{Kopeliovich:2002} except that our ansatz 
has to be 
piecewise linear to be able to generate configurations with a huge number 
of layers. After defining the ansatz
for an arbitrary number of layers, we will show that  
for a single layer configuration the ansatz produces configurations that
are in good agreements with the rational map ansatz configuration. 
Then we will use the new ansatz to construct configurations that are made 
out of a very large no of layers. 

We have shown, in the previous section, that one can construct shell like 
structures with very large Baryon numbers.
At large baryon numbers, the 
Skyrmions resemble shell like structures. That is, the fields are constant 
nearly everywhere except in a small region corresponding to the shell.
In that region, the profile look like linear functions smoothly linked to the
constant parts at the edges (cfr. Fig \ref{Num1step} ). 
Motivated by this we approximate the fields by the ramp-functions
\begin{align}
F(x)&=\frac{N \pi}{2} - (x-x_{0})\frac{\pi}{W},\qquad \qquad &(x_0 - NW/2)
\leq x \leq(x_0 + NW/2) \\
\mu(x)&=\frac{M}{2} + (x-x_{0})\frac{M}{N W},\qquad \qquad &(x_0 - NW/2)
\leq x \leq(x_0 + NW/2) \\
A(x)&=\frac{(1+A_{0})}{2} + (x-x_{0})\frac{(1-A_{0})}{N W},\qquad \qquad &
(x_0 - NW/2)\leq x \leq(x_0 + NW/2) 
\end{align}

In the above there are four free parameters, namely the central radius $x_0$ 
of the shell over which the fields change, the width of the shell $W$, the 
mass field at spatial infinity $M$ and the value of the metric field at the 
origin $A_0$ such that $\lim_{x\rightarrow \infty} = 0$. $N$ is 
the number of layers we wish to study and, as such, is treated as an input 
parameter.

The picture is of a gravitating skyrmion with very high baryon number existing 
as $N$ thin layers or shell of small thickness.

The above ansatz, allow us to find an approximation to the integrated energy. 
To do this we use the fact that the shell width is much smaller than the 
radius at large baryon numbers. In particular to evaluate the action integral
we can approximate expressions of the type $\int G(x) \sin^p F(x) $ for any 
function $G(x)$ that varies very little over the width of the shell by 
$\int  G(x_0) \sin^p F(x)$. We then use the fact that
\begin{equation}
\int_{x_0-NW/2}^{x_0+NW/2} \sin^p F(x) = 
\frac{N \pi}{W}\,\int_{0}^{\pi} \sin^p y \,dy.
\end{equation}

This leads to the following expression for the energy:
\begin{eqnarray}
\label{ShellE}
E &=& -\frac{16 \pi F_{\pi}}{e} 
\bigg[\frac{M}{\alpha}\left({1+A_0 \over 2}\right)- {\pi^2\over W}
  \bigg[\left({1+A_0 \over 2}\right)
        \left(x_0^2-M x_0 + {W^2 \over 12} - { M W \over 6}\right)
\nonumber\\
  &+&\left({1-A_0 \over W}\right) 
    \left({W^2 x_0\over 6} - {M W^2 \over 12}- {M W x_0 \over 6}
    \right)
  \bigg]
\nonumber\\
  &+& \mathcal{B} \left({1+A_0 \over 2}\right) \frac{1}{2}
   \left({M \pi^2 \over W x_0} - W - { \pi^2 \over W}\right)
  - {3 \mathcal{I} W \over 16 x_0^2} \left({1+A_0 \over 2}\right) 
  - {M\over \alpha}\bigg]
\end{eqnarray}
To find the configurations which minimize this energy we first 
minimised it with respect to $A_0$ and $M$ algebraically in order to find an 
expression for the energy as a function of the width and radius only. Then 
we minimised this numerically using \textit {Mathematica}.
We will now discuss the features of these configurations.

First of all, we must compare the results obtained with the ramp-profile
when $N=1$ and compare them to the result obtained with the full profile.
Tables. \ref{1shellApp} and \ref{2shellApp} show the properties of solutions 
we obtained using the ramp-profile approximation, again at 
$\alpha = 1 \times 10^{-6}$. 
All the general features of the full numerical solutions are reproduced. 
In particular, the approximate $\mathcal{B}^{\frac{1}{2}}$ scaling and then 
decrease of the radius, the decreasing ADM mass and the differences between 
the double and single layer solutions are all exhibited by the data obtained 
using the ramp-profile approximation.

\begin{table}
\begin{center}
\begin{tabular}{|c|c|c|c|c|}
\hline
$\mathcal{B}$ & $R (\frac{2}{F_{\pi}{e}})$ & $W$ & 
$M_{ADM} (\frac{F_{\pi}}{2 e}) $ & $S_{min}$ \\
\hline
100 & 8.3063 & 3.1286 & 1.1023 & 0.9999 \\
\hline
500 & 18.6031 & 3.1386 & 1.1160 & 0.9997 \\
\hline
$1 \times 10^3$ & 26.313 & 3.1397 & 1.1195 & 0.9996 \\
\hline
$1 \times 10^4$ & 83.206 & 3.1396 & 1.1254 & 0.9987 \\
\hline
$1 \times 10^5$ & 262.94 & 3.1357 & 1.1266 & 0.9960 \\
\hline
$1 \times 10^6$ & 829.60 & 3.1230 & 1.1251 & 0.9872 \\
\hline
$1 \times 10^7$ & 2604.2 & 3.0825 & 1.1186 & 0.9595 \\
\hline
$1 \times 10^8$ & 8032.8 & 2.9512 & 1.0972 & 0.8713 \\
\hline
$1 \times 10^9$ & 22899 & 2.4837 & 1.0272 & 0.5772 \\
\hline
$2 \times 10^{9}$ & 29121 & 2.1092 & 0.9818 & 0.3645 \\
\hline
$2.8 \times 10^{9}$ & 29098 & 1.6623 & 0.9505 & 0.1380 \\
\hline
$2.83 \times 10^{9}$ & 28514 & 1.6066 & 0.9495 & 0.1119 \\
\hline
$2.839 \times 10^{9}$ & 28024 & 1.5671 & 0.94922 & 0.09373 \\
\hline
$2.8397 \times 10^9$ & 27869.3 & 1.5556 & 0.94924 & 0.08845 \\
\hline
$2.83975 \times 10^9$ & 27869.8 & 1.5524 & 0.94925 & 0.08699 \\
\hline
$2.839752 \times 10^9$ & 27822 & 1.5521 & 0.94925 & 0.08687 \\
\hline
\end{tabular}
\end{center}
\caption{\emph{Table of properties of the single layer step ansatz 
configurations for varying the baryon number at fixed 
$\alpha=1 \times 10^{-6}$}}
\label{1shellApp}
\end{table}

\begin{table}
\begin{center}
\begin{tabular}{|c|c|c|c|c|}
\hline
$\mathcal{B}$ & $R (\frac{2}{F_{\pi}{e}})$ & $W$ & 
$M_{ADM} (\frac{F_{\pi}}{2 e}) $ & $S_{min}$ \\
100 & 5.7924 & 3.0428 & 1.0692 & 0.9999 \\
\hline
$1 \times 10^3$ & 18.5788 & 3.1305 & 1.1047 & 0.9995 \\
\hline
$1 \times 10^4$ & 58.8202 & 3.1380 & 1.1201 & 0.9983 \\
\hline
$1 \times 10^5$ & 185.8420 & 3.1332 & 1.1246 & 0.9944 \\
\hline 
$1 \times 10^6$ & 585.7950 & 3.1153 & 1.1233 & 0.9820 \\
\hline
$1 \times 10^7$ & 1833.0500 & 3.0578 & 1.1143 & 0.9428 \\
\hline
$1 \times 10^8$ & 5587.3600 & 2.8688 & 1.0840 & 0.8172 \\
\hline
$9 \times 10^8$ & 14147.1782 & 2.1859 & 0.9900 & 0.4065 \\
\hline
$9.764724 \times 10^8$ & 14472.3851 & 2.1276 & 0.9837 & 0.3746 \\
\hline
$1 \times 10^9$ & 14560.5000 & 2.1092 & 0.9818 & 0.3646 \\
\hline
$1.4 \times 10^9$ & 14549.0000 & 1.6623 & 0.9505 & 0.1381 \\
\hline
$1.41963 \times 10^9$ & 13994.0523 & 1.5644 & 0.9492 & 0.0926 \\
\hline
$1.419635134 \times 10^9$ & 13993.2000 & 1.5643 & 0.9492 & 0.0925 \\
\hline
\end{tabular}
\end{center}
\caption{\emph{Table of properties of the double layer step ansatz 
configurations for varying the baryon number at fixed 
$\alpha=1 \times 10^{-6}$}}
\label{2shellApp}
\end{table}

Quantitatively though, there are some differences. The approximation allows a 
significant increase in the maximum allowed baryon charge. Also, the radius 
of configurations obtained using the approximation, tend to be smaller than 
those obtained numerically. If we concentrate on the baryon numbers greater 
than $10^5$ so as to ensure our approximation, that the width is much smaller 
than the radius, is valid, then at worst we find a discrepancy in the ADM 
mass of $11 \%$ and in the radius of $7 \%$.

In general then, the data seems to confirm the reliability of the ramp-profile 
approximation. In fact the approach will be even more reliable at the 
extremely high values of the baryon number that we are interested in. This is 
because the radius of solutions is of orders of magnitudes greater than the 
width in such a regime, consistent with the approximations we have made.

Moreover, whilst searching for minima of the energy does not allow us to probe 
both branches of solutions, it does allow us to locate the value of 
$\alpha_{crit}$. We again obtain the approximate trend 
$\alpha_{crit} \propto \mathcal{B}^{- \frac{1}{2}}$, for large $\mathcal{B}$. 


Now in order to say anything about the possibility of baryon stars in the 
Skyrme model we need to be able to verify that the decrease in the ADM mass 
per baryon we observed at low and moderate baryon numbers, extends to baryon 
numbers of order $10^{58}$ for $\alpha = 7.3 \times 10^{-40}$.

Table. \ref{realApp} summarizes our solutions in such a regime. 
Firstly we consider constructing a single layer self-gravitating Skyrmion 
with these values. We do indeed see that the configuration is bound. This 
is verified by checking that the ADM mass is lower (even at this significantly 
lower value of $\alpha$) than for the $\mathcal{B}=1$ hedgehog. 
So the possibility of baryon stars in the Einstein-Skyrme 
model cannot be ruled out on the grounds of energy.

The Skyrmion exists as a giant thin shell, and the large 
baryon charge is distributed as a tight lattice over this. However a hollow 
shell is clearly not a realistic construction for a 
neutron star. This fact manifests itself in the extremely high radius of the 
configuration. Transferring to standard units, the single layer 
$\mathcal{B}=10^{58}$ gravitating Skyrmion has a radius of 
$2.42 \times 10^{10} km$ !

To address this issue, we can use a large number of layered Skyrmions as
discussed earlier. This has several benefits. Firstly, as we are distributing 
the baryon number through a larger volume, then at a given baryon density 
the necessary radius can decrease. Similar to what we see in the double layer 
results. On top of this, we expect the radius to decrease further due to extra 
gravitational compression, as the outer layers of the Skyrmions feel the 
attraction of inner layers. Finally, the many layer approach is also a more 
realistic construction of a solid baryon star.

The results for using more and more layers in the construction (for fixed 
$\mathcal{B}$ and $\alpha$), are also presented in Fig. \ref{realApp}. We 
note that not only does the radius decrease significantly, but the added 
gravitational binding further improves the energies of the configurations, 
reflected in the low ADM masses obtained. There appears to be a critical 
number of layers that can be used before there ceases to be any solutions and 
although the value of $S_{min}$ is close to zero at this point, the star 
still has not collapsed to form a black hole. Finally, we note that the 
radius of the Skyrmion at the critical number of layers is approximately 
$20.91 km$. This is comparable to a real neutron star, with a typical radius 
of $10 km$.

We reemphasise here that our approach to embedding shells of baryons is quite 
crude. For few shells and large baryon number, we might reasonably believe 
that baryon number does not chance significantly from one shell to the next. 
However, when we embed many shells we should really consider that the baryon 
number of the inner most shells would likely be significantly less than the 
that of the outer shells. Nevertheless, our naive embedding has produced some 
interesting properties. In a future work we hope to improve our 
multi-layer construction to obtain a more realistic description of a 
baryon star.


\begin{table}
\begin{center}
\begin{tabular}{|c|c|c|c|c|}
\hline
$N_{Shell}$ & $R (\frac{2}{F_{\pi}{e}})$ & $W$ & $M_{ADM}/(6 \pi^2 B)$ & $S_{min}$ \\
\hline
$1 \times 10^2$ & $8.3236 \times 10^{27}$ & 3.1416 & 1.1285 & 1.0000 \\
\hline
$1 \times 10^3$ & $2.6321 \times 10^{27}$ & 3.1416 & 1.1285 & 1.0000 \\
\hline
$1 \times 10^4$ & $8.3236 \times 10^{26}$ & 3.1416 & 1.1285 & 1.0000 \\
\hline
$1 \times 10^5$ &  $2.6321 \times 10^{26}$ & 3.1416 & 1.1285 & 1.0000 \\
\hline
$1 \times 10^6$ & $8.3236 \times 10^{25}$ & 3.1416 & 1.1285 & 1.0000 \\
\hline
$1 \times 10^7$ & $2.6321 \times 10^{25}$ & 3.1416 & 1.1285 & 1.0000 \\
\hline
$1 \times 10^8$ & $8.3236 \times 10^{24}$ & 3.1416 & 1.1285 & 1.0000 \\
\hline
$1 \times 10^9$ & $2.6321 \times 10^{24}$ & 3.1415 & 1.1285 & 1.0000 \\
\hline
$1 \times 10^{10}$ & $8.3234 \times 10^{23}$ & 3.1415 & 1.1285 & 0.9999 \\
\hline
$1 \times 10^{11}$ & $2.6319 \times 10^{23}$ & 3.1412 & 1.1285 & 0.9997 \\
\hline
$1 \times 10^{12}$ & $8.3216 \times 10^{22}$ & 3.1402 & 1.1283 & 0.9991 \\
\hline
$1 \times 10^{13}$ & $2.6301 \times 10^{22}$ & 3.1373 & 1.1278 & 0.9971 \\
\hline
$1 \times 10^{14}$ & $8.3034 \times 10^{21}$ & 3.1280 & 1.1263 & 0.9907 \\
\hline
$1 \times 10^{15}$ & $2.6118 \times 10^{21}$ & 3.0986 & 1.1213 & 0.9705 \\
\hline
$1 \times 10^{16}$ & $8.1147 \times 10^{20}$ & 3.0037 & 1.1057 & 0.9063 \\
\hline
$1 \times 10^{17}$ & $2.4001 \times 10^{20}$ & 2.6810 & 1.0552 & 0.6977 \\
\hline
$5 \times 10^{17}$ & $8.2066 \times 10^{19}$ & 1.7888 & 0.9552 & 0.2036 \\
\hline
$5.3 \times 10^{17}$ & $7.4172 \times 10^{19}$ & 1.6227 & 0.9491 & 0.1247 \\
\hline
$5.33 \times 10^{17}$ & $7.1871 \times 10^{19}$ & 1.5625 & 0.94866 & 0.0971 \\
\hline
$5.3306 \times 10^{17}$ & $7.1597 \times 10^{19}$ & 1.5549 & 0.94868 & 0.0936 \\
\hline
$5.33065 \times 10^{17}$ & $7.1525 \times 10^{19}$ & 1.5528 & 0.948694 & 0.0927 \\
\hline
$5.330657 \times 10^{17}$ & $7.1506 \times 10^{19}$ & 1.5523 & 0.948692 & 0.0924\\
\hline
\end{tabular}
\caption{\emph{Table of properties of the step ansatz configurations for 
varying the number of embedded shells at fixed $\mathcal{B}=10^{58}$ and 
$\alpha=7.3 \times 10^{-40}$.}}
\label{realApp}
\end{center}
\end{table}

\newpage
\section{Conclusions}

Previous work on the Einstein-Skyrme model highlighted a considerable problem
with using the Skyrmions as a model for baryon stars. Namely, multibaryon 
hedgehog Skyrmions were simply not energetically favourable states. We have 
shown that this is simply a consequence of a poor ansatz for the true Skyrmion 
and, having used the more appropriate rational map ansatz, we have generated 
energetically favourable configurations of multibaryons. 

We also observe the interesting property that near the critical coupling, the 
Skyrmions can decrease in radius as we add more baryons. This hints towards 
the similar behaviour exhibited by real neutron stars.

Although the rational map ansatz does not have an exact radial symmetry, 
at large scale it does. The anisotropy only appears at the nucleon scale.

Finally, since we started with the motivation of studying baryon stars within 
the Skyrme model, it is interesting to compare the features of our 
configurations with those of neutron stars. For realistic values, 
$\mathcal{B}=10^{58}$ and $\alpha=7.3 \times 10^{-40}$ we find a minimal 
energy single layer configuration with radius=$2.42 \times 10^{10}km$. This 
is clearly too large for a neutron star (which is of order 10km. in radius). 
This is to be expected however due to the shell model we have taken. Firstly, 
as we are distributing the baryons over the surface area rather than 
throughout the volume of the star we naturally must require a much larger star 
for a given baryon number. This effect is two-fold in that if we were 
distributing the baryons throughout the volume, outer layers would feel the 
attraction of inner layers and enhanced radial compression would occur. The 
loss of such an effect is pronounced when we are considering realistically 
small values of the coupling.

It seems therefore that the way to construct baryon stars in the Skyrme model 
is to consider embedding shells of baryons within shells. This gives 
rise to more appropriate specifications for the star and is also more 
realistic. We do indeed observe such improvements for a many layered 
configuration. In fact the radius of $\mathcal{B}=10^{58}$ gravitating 
Skyrmion (at realistic $\alpha$), can be decreased in this manner to 
approximately $20.91 km$.

We note however that this approach to shell embedding has only be done 
naively thus far. We have only considered the case where the baryon number 
is equal for each shell. We really should allow the baryon number(and hence 
the rational map quantities) to vary over the shells. One approach towards 
this would be to assume that the baryon density is a constant over the shells. 
An even better approach would be to allow this to be a smoothly varying 
function that must be determined by minimising the energy. This will give 
a more realistic description of baryon stars within the Einstein-Skyrme model, 
as traditional descriptions of neutron stars also involve many strata, of 
differing neutron density. We are currently investigating such 
configurations.  

\section{Acknowledgements}
GIP is supported by a PPARC studentship.


\begin{thebibliography}{99}

\bibitem{Bizon:1992}
P. Bizon \& T. Chmaj
``Gravitating Skyrmions''
Phys. Lett. B {\bf 297} (1992), 55-62

\bibitem{Skyrme:1961}
T. H. R. Skyrme,
``A Non-Linear Field Theory''
Proc. Roy. Soc. A {\bf 260} (1961), 127-138

\bibitem{Skyrme:1962}
T. H. R. Skyrme,
``A Unified Theory of Mesons and Baryons''
Nucl. Phys.{\bf 31} (1962)



\bibitem{Witten:1983}
E. Witten
``Global Aspects of Current Algebra''
Nucl. Phys. {\bf B223} (1983), 422-433

\bibitem{Glendenning:1988}
N. K. Glendenning, T. Kodama \& F. R. Klinkhamer
``Skyrme Topological Soliton Coupled to Gravity''
Phys. Rev. D {\bf 38} Number 10 (1988), 3226-3230

\bibitem{Volkov:1999}
M. S. Volkov \& D. V. Gal'tsov
``Gravitating Non-Abelian solitons and Black Holes with Yang-Mills Fields''
Physics Reports, {\bf 319}, Numbers 1-2, 1-83 (1999)

\bibitem{Luckock:1986}
H. Luckock \& I. Moss
``Black Holes HAve Skyrmion Hair''
Phys. Lett {\bf B176} (1986),341-345

\bibitem{Droz:1991}
S. Droz, M. Heusler \& N. Straumann
``New Black Hole Solutions with Hair''
Phys. Lett. {\bf B268} (1991), 371-376

\bibitem{BS1997}
R.A. Battye, P.M. Sutcliffe
`` MULTI - SOLITON DYNAMICS IN THE SKYRME MODEL.''
Phys.Lett. {\bf B391} (1997), 150-156

\bibitem{Houghton:1998}
C. Houghton, N. Manton \& P. Sutcliffe
``Rational MAps, Monopoles and Skyrmions''
Nucl. Phys. {\bf B510} (1998), 507-537

\bibitem{Battye:2001}
R. A. Battye \& P. M. Sutcliffe
``Skyrmions, Fullerenes and Rational Maps''
Rev. Math. Phys. {\bf 14} (2002), 29-86

\bibitem{Manton:2000}
N. S. Manton \& B. M. A. G. Piette
``Understanding Skyrmions using Rational Maps''
hep-th/0008110
Understanding Skyrmions Using Rational Maps: 
Proceedings of the European Congress of Mathematics, Barcelona
2000, eds. C.Casacuberta et al., Progress in Mathematics, Birkhauser, Basel 
{\bf Vol. 201} (2001) 469-479

\bibitem{Kopeliovich:2001}
V. B. Kopeliovich
``The Bubbles of Matter from MultiSkyrmions''
JETP Lett. {\bf 73} (2001), 587-591; Pisma Zh.Eksp.Teor.Fiz. 73 (2001), 667-671

\bibitem{Kopeliovich:2002}
V. B. Kopeliovich
``MultiSkyrmions and Baryonic Bags''
J.Phys. {\bf G28} (2002), 103-120
\end{thebibliography}
\end{document}